%% file: JCIN22.tex
\documentclass[lettersize,journal]{IEEEtran}
\usepackage{xcolor}
\usepackage{makecell}
\usepackage{multirow,graphicx}
\usepackage[colorlinks,linkcolor=blue,urlcolor=blue,citecolor=blue]{hyperref}

\begin{document}


\title{WirelessLLM: Empowering Large Language Models Towards Wireless Intelligence}

\author{Jiawei~Shao, Jingwen~Tong, Qiong~Wu, Wei~Guo, Zijian~Li, Zehong~Lin, and~Jun~Zhang,~\IEEEmembership{Fellow,~IEEE}
\thanks{This work was supported by the Hong Kong Research Grants Council under the Areas of Excellence scheme grant AoE/E-601/22-R and NSFC/RGC Collaborative Research Scheme grant CRS\_HKUST603/22.

The authors are with the Department of Electronic and Computer Engineering, Hong Kong University of Science and Technology, Hong Kong (E-mail: jiawei.shao@connect.ust.hk, \{eejwentong, eeqiongwu, eeweiguo\}@ust.hk, zijian.li@connect.ust.hk, \{eezhlin, eejzhang\}@ust.hk).
The corresponding author is J. Zhang.}}


\maketitle

{\bf\textit{Abstract---}
The rapid evolution of wireless technologies and the growing complexity of network infrastructures necessitate a paradigm shift in how communication networks are designed, configured, and managed. Recent advancements in Large Language Models (LLMs) have sparked interest in their potential to revolutionize wireless communication systems. 
However, existing studies on LLMs for wireless systems are limited to a direct application for telecom language understanding.
To empower LLMs with knowledge and expertise in the wireless domain, this paper proposes WirelessLLM, a comprehensive framework for adapting and enhancing LLMs to address the unique challenges and requirements of wireless communication networks.
We first identify three foundational principles that underpin WirelessLLM: knowledge alignment, knowledge fusion, and knowledge evolution.
Then, we investigate the enabling technologies to build WirelessLLM, including prompt engineering, retrieval augmented generation, tool usage, multi-modal pre-training, and domain-specific fine-tuning.
Moreover, we present three case studies to demonstrate the practical applicability and benefits of WirelessLLM for solving typical problems in wireless networks.
Finally, we conclude this paper by highlighting key challenges and outlining potential avenues for future research.
\\[-1.5mm]

}.


\section{Introduction}

\input{1.Intro} 

\input{2.Overview_of_LLMs} 

\input{3.WirelessLLM}

\input{4.Enabling_tech}

\section{Case Studies}\label{sec:Case_Studies}
In this section, we compare WirelessLLM with other methods across three different tasks is the testing wireless environments using the same settings. Specifically, we demonstrate the use of WirelessLLM to address the tasks of power allocation, spectrum sensing, and protocol understanding in wireless networks by employing the techniques of tool manipulation, prompt engineering, and RAG, respectively.
\input{case_study_wei}

\input{case_study_jingwen}

\input{case_study_qiong}

\input{6.Challenges}

\section{Conclusion}
\label{sec:conclusion}

This paper developed a novel framework, WirelessLLM, to adapt and enhance LLMs specifically for wireless communication systems. 
This adaptation is crucial to overcome their inherent limitations and effectively address the challenges posed by the wireless domain.
Specifically, we identified three core principles that underpin WirelessLLM and investigated the enabling techniques.
Our case studies demonstrated the effectiveness of WirelessLLM, showcasing its superior performance in solving complex wireless communication problems compared to generic LLMs.
Moreover, we discussed the key challenges and open research directions in enhancing the efficiency, scalability, and security of WirelessLLM. 
The clear potential of LLMs to revolutionize the field of wireless communications is evident.
As we continue to refine and expand the capabilities of WirelessLLM, we anticipate it will become a fundamental tool in the design and management of the next generation of wireless systems.

\bibliographystyle{IEEEbib}
\bibliography{refs}

\end{document}

%% file: 1.Intro.tex
With the increasing complexity and diversity of service requirements, future wireless systems are expected to leverage artificial intelligence (AI) algorithms to design and optimize wireless networks across different layers \cite{shi2020communication,letaief2019roadmap,xu2024large,shao2020communication}.
Most traditional approaches are based on optimization theory techniques, which only work if appropriate mathematical models are available.
As wireless networks become increasingly complex and heterogeneous, formulating such models is more and more difficult.
In contrast, deep learning (DL) approaches can adapt to different scenarios by directly learning from data, without the need for explicit mathematical modeling \cite{zhu2020toward,shen2022graph,shao2021learning,ShaoTWC,he2018deep,shao2023task}. 

Recently, key AI domains have undergone a paradigm shift with the emergence of large language models (LLMs) \cite{radford2019language,brown2020language,touvron2023llama}.
These models, trained on vast amounts of diverse text data in an autoregressive manner, have achieved remarkable performance in natural language processing (NLP).
Notably, they possess the ability to capture universal knowledge and demonstrate significant adaptability to various downstream tasks.
The profound competencies have enabled domain-adapted LLMs to successfully address complex challenges across diverse fields, including robot embodied intelligence \cite{fan2024embodied}, chip design \cite{liu2023chipnemo}, and protein structure estimation \cite{ferruz2022controllable}.
Given the expectation that future wireless systems will require ever more sophisticated and adaptable technologies to manage increasing complexity and user demands, a promising research direction is to explore LLMs in the context of automating and optimizing network operations and enhancing user experience.

\begin{figure*}
    \centering
    \includegraphics[width = 0.99\linewidth]{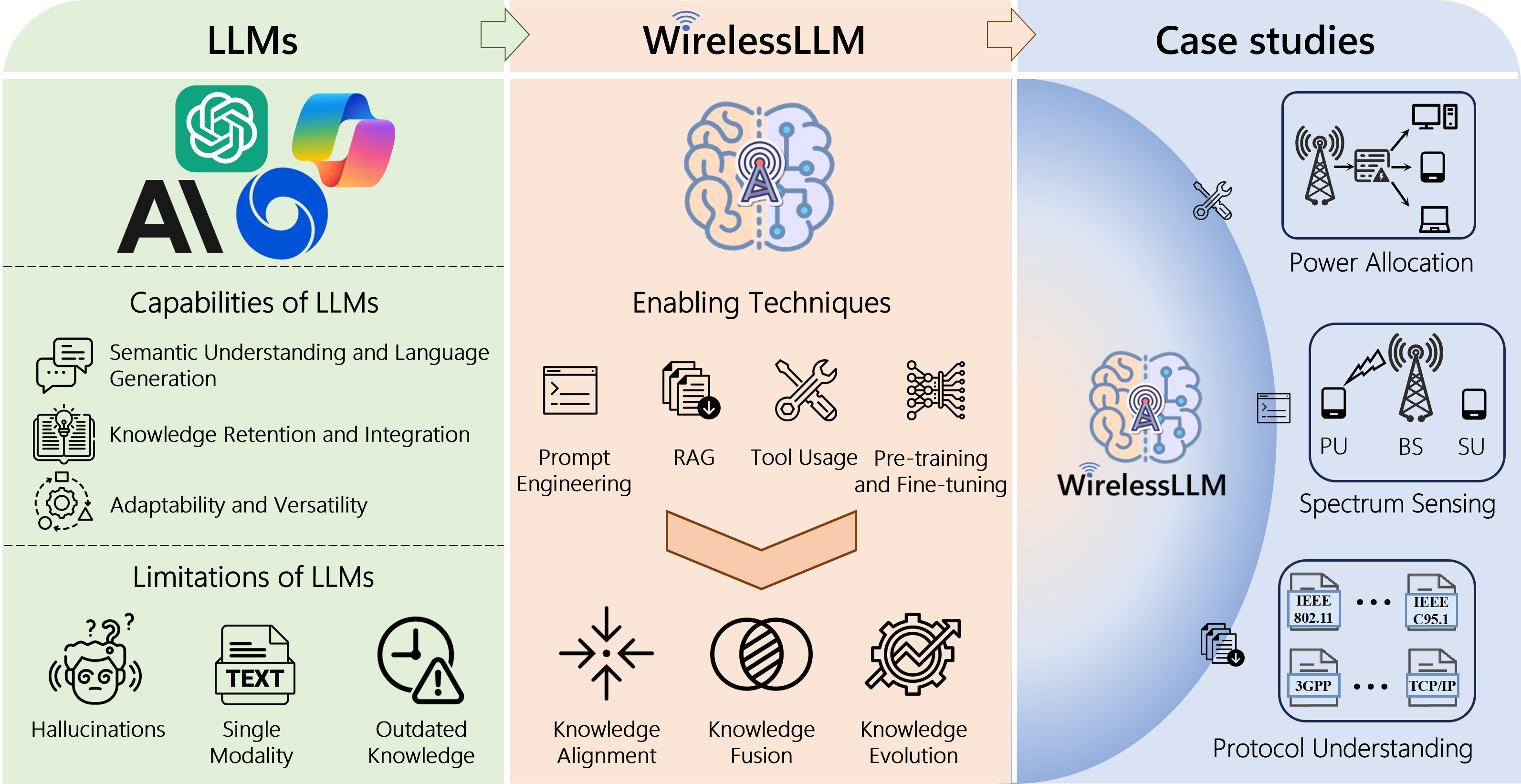}
    \caption{The overview of WirelessLLM.}
    \label{fig:overview}
\end{figure*}

Recent efforts have utilized LLMs for semantic communication \cite{xie2024towards}, information synthesis \cite{kotaru2023adapting}, network optimization \cite{tarkoma2023ai}, and edge inference \cite{shen2024large}.
However, since the LLMs are only trained on text corpora, these models cannot capture the multi-modal data generated by the diverse sensors and interfaces of wireless systems. Consequently, their role is restricted to that of chatbots, rather than serving as a comprehensive solution. 
Besides, a significant issue of LLMs is their tendency to hallucinate by producing outputs that mimic human responses but lack factual grounding.
Moreover, these models lack an understanding of the complex physics that govern the wireless environment.
For example, an LLM may suggest power allocation strategies that violate fundamental constraints, such as the maximum transmit power, or that cause severe interference.
In addition, as wireless systems evolve to incorporate new protocols and standards, continuous adaptation of these models becomes essential to ensure their practicality and effectiveness in real-world scenarios.

Therefore, it is important to empower LLMs with domain-specific knowledge to effectively adapt them for wireless systems.
This includes aligning their output with the characteristics and constraints of wireless environments, equipping them with the capability to understand multi-modal data, and continuously evolving their knowledge with the rapid advancements in wireless technologies and dynamic channel states.
This paper aims to unlock the potential of LLMs in wireless communications. Our main contributions are summarized as follows:
\begin{itemize}
    \item We propose WirelessLLM, a framework that empowers LLMs with knowledge and expertise in the wireless domain
    to facilitate their adaptation to wireless systems.
    The framework is underpined by three foundational principles: knowledge alignment, knowledge fusion, and knowledge evolution.
    \item We delve into a suite  of enabling techniques that facilitate the deployment of WirelessLLM, including prompt engineering, retrieval augmented generation, tool usage, multi-modal pre-training, and domain-specific fine-tuning.
    \item We present three case studies to showcase the effectiveness of WirelessLLM in practical wireless communication scenarios, including power allocation, spectrum sensing, and protocol understanding. Proof-of-concept results demonstrate that incorporating domain-specific knowledge significantly improves the response quality of WirelessLLM compared to LLMs without wireless context.
    \item We highlight the key challenges in constructing WirelessLLM, covering aspects such as dataset collection for model training, computational complexity in inference, and security concerns related to wireless systems.
    In addition, we discuss potential approaches and future research directions to mitigate these bottlenecks, thereby facilitating the efficient implementation of WirelessLLM.
\end{itemize}

Figure \ref{fig:overview} illustrates an overview of WirelessLLM.
The remainder of this paper is organized as follows. In Section 2, we present an overview of LLMs.
Section \ref{sec:WirelessLLM} proposes the WirelessLLM framework, and Section \ref{sec:Enabling_Tech} investigates the enabling techniques.
Next, Section \ref{sec:Case_Studies} presents three case studies to demonstrate the effectiveness of WirelessLLM.
Finally, Section \ref{sec:challenges} points out existing challenges and future research directions, and Section \ref{sec:conclusion} concludes the paper. 



%% file: 2.Overview_of_LLMs.tex
\section{Overview of LLMs}\label{sec:Overview}

To explore the potential of LLMs for wireless networks, it is essential to begin by establishing a foundational understanding.
This section delves into the core concepts and principles of LLMs, explores their key capabilities, and examines their limitations.

\subsection{Fundamentals of LLMs}

LLMs represent a significant evolution in the field of NLP. 
These models, built on the Transformer architecture introduced by Vaswani et al. \cite{vaswani2017attention}, have transformed the way machines understand and generate human language by leveraging deep neural networks trained on vast amounts of text data.
The core innovation driving this success  is the attention mechanism. Rather than processing text in a sequential fashion like Long Short-Term Memory networks (LSTMs) \cite{sundermeyer2012lstm_LSTM} and Recurrent Neural Networks (RNNs) \cite{speech_graves2013speech}, the attention mechanism allows the model to process tokens by considering the relationship between each token and all other tokens in a sentence. This enables the model to effectively handle long-range dependencies. 
By assigning weights to input tokens based on their importance, the attention mechanism empowers the model to prioritize relevant information. 
More specifically, this mechanism first calculates query, key, and value vectors for each token in the input sequences. 
Then, it computes attention scores by taking the dot product of the query vector of one token with the key vectors of all other tokens. 
These scores are used to weight the value vectors, allowing the model to learn complex interactions and gain a comprehensive understanding of the input.
In addition, LLMs benefit from the scalability of the Transformer architecture.
Recent studies in \cite{kaplan2020scaling,hoffmann2022training} have shown that increasing the number of parameters in a model correlates with improved performance across various NLP tasks. 

Building an LLM typically involves a pre-training phase and a fine-tuning phase.
During the pre-training phase, the model learns general language patterns from a vast dataset of diverse text sources, encompassing literature, web content, scientific articles, and others.
This objective is achieved primarily through self-supervised learning techniques, such as masked language modeling (MLM) \cite{salazar2020masked} and autoregressive language modeling (ALM) \cite{yang2019xlnet}.
MLM enforces the model to learn to predict hidden words in a sentence, while ALM involves predicting the next word in a sequence based on the preceding context.
After pre-training on an extensive corpus, the LLM captures a wide range of linguistic information, including syntactic structures, semantic relationships, and contextual understanding.
Subsequently, the model undergoes a fine-tuning phase \cite{ding2023parameter}, where it is further trained on a more specific dataset tailored to target tasks or domains. 
This process allows the model to adapt its responses to be more relevant and accurate for its intended application, enhancing its performance in specialized domains.

\subsection{Capabilities of LLMs}

LLMs have demonstrated remarkable capabilities that revolutionize various aspects of NLP and beyond. 
The following content discusses three distinct capabilities:

\textbf{Semantic understanding and language generation.} 
LLMs tokenize raw text data into a sequence of integers and convert them into dense vector embeddings.
These embeddings capture an intrinsic understanding of the input text, including contextual relationships and semantic nuances, which are then processed through multiple layers of neural networks to extract higher-level features and patterns.
By utilizing these detailed representations, LLMs can achieve deep semantic understanding and generate syntactically coherent text.
Specifically, they can parse and comprehend intricate sentence structures, disambiguate word meanings, and produce human-like language that adheres to grammatical rules and maintains contextual coherence.

\textbf{Knowledge retention and integration.} 
LLMs have the ability to retrieve vast amounts of knowledge from the training data and store it within model parameters, effectively creating a knowledge base.
This includes factual knowledge, linguistic patterns, and cultural nuances.
When given a query, LLMs can effectively select relevant pieces of information, synthesize them in a meaningful way, and present the results coherently and fluently.
In addition to the direct integration of knowledge, LLMs exhibit the ability to solve complex tasks by decomposing them into simpler components and addressing each part sequentially \cite{wei2022chain}.
Such multi-step reasoning enables LLMs to utilize intermediate results for deriving the final answer, demonstrating their logical thinking and problem-solving skills.

\textbf{Adaptability and versatility.} 
After the initial pre-training phase, LLMs can acquire the general abilities for solving various tasks.
With minimal task-specific fine-tuning or even simple prompting, these models have proven to be highly adaptable and versatile, making them proficient in diverse domains.
For instance, a study \cite{zheng2023survey} has shown that LLMs are valuable tools in assisting developers with code generation.
In the medical field, LLMs have demonstrated the capability to parse medical literature and aid in diagnosing symptoms described in natural language \cite{thirunavukarasu2023large}.
More recently, the authors of \cite{maatouk2023teleqna, bariah2023understanding} have showcased the ability of LLMs to understand telecommunication lexicon and standards.

\subsection{Limitations of LLMs}

Despite the impressive capabilities of LLM, they also face several limitations that can affect their performance and potential, including:

\textbf{Hallucinations.} 
One major concern is that LLMs can generate plausible but factually incorrect or nonsensical information \cite{li2023halueval}.
This is because they generate responses based on statistical patterns of the training data rather than retrieving facts from reliable sources.
This tendency to hallucinate can mislead users or degrade the trustworthiness of LLMs, especially in critical applications. 
Moreover, LLMs may exhibit misalignment between their outputs and the intended goals or values of their users, resulting in responses that are inappropriate, biased, or out of context~\cite{kopf2024openassistant}.

\textbf{Single modality.} 
Most LLMs are primarily trained on textual data and therefore struggle to directly process and understand other modalities such as images, audio, and wireless physical signals.
While some progress \cite{shen2024hugginggpt,wang2024survey} has been made in transforming these non-textual data sources into text-based formats, these conversions often result in a loss of critical information and nuances inherent in the original data.
This restricts the applicability and effectiveness of LLMs in multidisciplinary contexts.

\textbf{Outdated knowledge.} 
LLMs encode a large amount of world knowledge.
However, their knowledge is frozen at the point of their last training update, making the models limited by the training data at that time.
This static nature renders LLMs less effective in fields that require up-to-date knowledge, such as recent developments, evolving protocols, and dynamic environmental states.

%% file: 3.WirelessLLM.tex
\section{WirelessLLM: Adapting LLMs to Wireless Systems}\label{sec:WirelessLLM}

With an understanding of the fundamentals, capabilities, and limitations of LLMs, this section proposes WirelessLLM, a framework to empower LLMs for wireless systems.
The key concept is to integrate telecom knowledge into LLMs, which encompasses physical realities, network protocols, and mathematical models.
Three principles of WirelessLLM include: \textbf{knowledge alignment}, which ensures that the model's outputs are closely aligned with physical reality and practical constraints; \textbf{knowledge fusion}, which involves synthesizing wireless physical signals to form a coherent understanding of the electromagnetic environment; and \textbf{knowledge evolution}, which requires continuously updating and refining the model based on dynamic channel conditions and human feedback.

\subsection{Knowledge Alignment}

Aligning LLMs with human values is a critical concern to avoid hallucination and misinformation.
It is essential to ensure that these models understand and reflect ethical considerations and human norms  for their safe and beneficial integration into society.
In the context of WirelessLLM, the concept of knowledge alignment means ensuring that the outputs of WirelessLLM are closely aligned with the physical realities and practical constraints of the wireless domain.
Wireless systems are fundamentally governed by the laws of physics, particularly electromagnetism, and are subject to various practical constraints, including bandwidth limitations, signal interference, and power constraints.
Therefore, WirelessLLM should be capable of effectively capturing the relationships among various physical signal strengths, antenna gain, propagation loss, and noise figures.
This necessitates a deep integration of domain-specific expertise into the knowledge base of WirelessLLM.
Besides, WirelessLLM should be able to accurately understand mathematical formulas and conduct computations.
The ability to compute and predict outcomes based on complex algorithms and signal behavior is fundamental for optimizing wireless communications systems. 

\subsection{Knowledge Fusion}

WirelessLLM is designed to fuse diverse knowledge in the wireless domain.
Accurately sensing and modeling real-world wireless environments is crucial for effectively capturing and understanding the complex interplay of various factors that shape wireless network performance. 
This requires WirelessLLM to jointly utilize domain-specific expertise and integrate multi-modality sensor data.
This, in turn, enables a deeper comprehension of the intricate dynamics and characteristics.
However, LLMs, being primarily text-based models, struggle to handle multi-modal and heterogeneous data, which can range from raw signal measurements to high-level user interactions and network performance metrics.
Previous approaches have attempted to directly transform multi-modality data into discrete tokens that LLMs can process. Nonetheless, the tokenization process often leads to a loss of precision and fidelity, ultimately compromising the performance of wireless systems.
To overcome this limitation, one potential direction is to develop hybrid models that combine the strengths of LLMs with specialized neural network models.
These hybrid models are designed to handle specific data modalities, representing physical symbols as continuous embeddings.
By interleaving these embeddings with text token embeddings, LLMs can effectively fuse knowledge from various sources and modalities. This fusion of knowledge empowers LLMs to acquire a more comprehensive and accurate understanding of wireless environments.

\subsection{Knowledge Evolution}

Knowledge evolution in WirelessLLM refers to the continuous adaptation and refinement of the model to perform adaptive modulation, select suitable antennas, and choose optimal wireless channels.
Unlike traditional models that are typically trained on static datasets, WirelessLLM must adapt to the ever-changing wireless environments and incorporate new knowledge, where the inherently dynamic nature is caused by factors such as user mobility, obstructions, and interference.
The first step in implementing knowledge evolution necessitates the integration of feedback loops.
These loops enable WirelessLLM to dynamically adjust its output based on real-time wireless environment conditions and user feedback.
By monitoring the performance metrics, such as signal-to-noise ratio, bit error rate, and user throughput, WirelessLLM should continuously update its parameters to enhance its decision-making performance.
Moreover, with the rapid development of wireless communication systems, new standards are continuously introduced, and user behavior can change accordingly.
To maintain the long-term effectiveness of WirelessLLM, it must evolve in parallel with these advancements and shifting user requirements. 
This requires WirelessLLM to continuously incorporate new knowledge about the latest technologies, protocols, and user behaviors.

%% file: 4.Enabling_tech.tex
\section{Enabling Techniques}\label{sec:Enabling_Tech}

In this section, we discuss the key techniques to build WirelessLLM, including prompt engineering, retrieval-augmented generation (RAG), tool usage, and the processes of pre-training and fine-tuning.

\subsection{Prompt Engineering}

Prompt engineering is a crucial method for enhancing the functionality of LLMs. 
This strategy designs specific task-oriented instructions, referred to as prompts, to empower LLMs to generate more accurate, relevant, and context-specific responses for downstream tasks.
The field of prompt engineering includes a range of techniques, from basic methods like zero-shot and few-shot prompting to more complex strategies such as chain-of-thought (CoT) prompting.
Zero-shot prompting relies on carefully crafted prompts that enable a model to address new tasks without any additional training.
Such prompts contain task descriptions, allowing the model to leverage its pre-trained knowledge to generate appropriate responses.
In contrast, few-shot prompting provides the model with a set of input-output pairs, which helps LLMs better understand the desired task and adapt their responses accordingly.
As demonstrated in \cite{brown2020language}, providing high-quality few-shot demonstrations improves the performance of LLMs compared with when no examples are provided.
In addition, CoT prompting encourages LLMs to break down a complex problem into simpler, more manageable steps, providing an explicit chain of reasoning that leads to the final answer.
In the context of applying LLMs to the domain of wireless systems, prompt engineering can be used to instruct WirelessLLM to prioritize domain-specific information for knowledge alignment. 
This can be achieved by crafting prompts that include specific details about wireless standards, protocols, or physical constraints.

\subsection{Retrieval-augmented Generation}

LLMs showcase impressive capabilities, but they also encounter challenges like hallucination, outdated knowledge, and bias in decision-making.
RAG is a promising solution by incorporating knowledge from external databases, which effectively bridges the gap between the specific domain context and the general knowledge base.
In wireless scenarios, external knowledge can encompass a wide range of sources, including device instruction manuals, algorithm textbooks, and wireless standards.
Once this information is retrieved, the generation component of RAG utilizes it to construct coherent and contextually relevant content, thereby enhancing the understanding of the relationships among various physical components in wireless systems.
For example, RAG can leverage external knowledge about modulation schemes and channel coding to perform more accurate link adaptation in response to different channel states.
Moreover, RAG with evolving knowledge functions as a sophisticated adaptive system. It incorporates a feedback mechanism that continuously refines its outputs based on new data, user interactions, and the latest protocols.
With this ever-evolving capability, RAG is capable of enhancing the accuracy and credibility of its generation, particularly for knowledge-intensive tasks, and allows for continuous integration of up-to-date information.

\subsection{Tool Usage}

LLMs are typically trained as text generators on massive plain text corpora, which limits their performance on tasks that are not best expressed in textual form.
To address this limitation, recent studies have explored the potential of using external plugins, software tools, and lightweight models to enhance their functionality beyond pure text understanding and generation.
WirelessLLM aims to integrate these tools into its framework to solve complex problems in wireless communications and provide easy-to-understand visual representations of complex results.
For example, WirelessLLM can utilize Matlab or Python functions as mathematical solvers to perform accurate calculations related to wave propagation, interference, and power allocation.
By incorporating specialized simulation tools and network modeling software, WirelessLLM enhances its ability to predict network performance, optimize configurations, and troubleshoot issues.
Similarly, WirelessLLM can leverage generative adversarial networks (GANs) \cite{yang2019generative, zou2023generative} and diffusion models \cite{sengupta2023generative, wu2023cddm} designed for channel generation to model dynamic wireless channels using limited data.
In addition, the incorporation of software packages such as Matplotlib and Seaborn within WirelessLLM enables it to visualize antenna patterns, network topologies, and frequency spectrum, thereby complementing the textual outputs.

\subsection{Multi-modal Pre-training and Fine-tuning}

The effectiveness of LLMs in wireless communications can be significantly enhanced by updating the model parameters on domain-specific datasets.
These data samples can be sourced from publicly available documents, generated through simulations, or collected from field tests.
Multi-modal pre-training involves optimizing models not only on text but also on other types of data relevant to the wireless domain, such as physical symbols, signal patterns, and environmental noise.
In general, the input of each modality is first extracted into feature embeddings by an independent encoder.
These features are then tokenized and concatenated as inputs for a Transformer architecture, which performs token-level matching based on the semantic correlation among different modalities.
By optimizing the cross-modal objective function, WirelessLLM is able to understand textual descriptions and analyze wireless signals to solve a wide range of tasks in the wireless communications domain.
Specifically, when multi-modal pre-training starts from a well-trained language model as tne base, the integration of additional modalities can be more seamlessly and quickly achieved.
Without having to retrain the language model from scratch, fine-tuning it with cross-modal data pairs allows WirelessLLM to adapt to the downstream wireless tasks.
The most widely used fine-tuning method is low-rank adaptation (LoRA), which freezes the pre-trained model weights and injects trainable rank decomposition matrices into each layer, greatly reducing the number of trainable parameters.
Moreover, to maintain the performance of WirelessLLM as wireless systems evolve, it is crucial to employ techniques such as reinforcement learning from human and channel feedback.
This mechanism ensures that WirelessLLM is continuously updated with the latest changes in wireless environments and user preferences.

%% file: case_study_wei.tex
\subsection{WirelessLLM for Power Allocation}\label{subsec:LLM4Opt}
\begin{figure*}[!t]
	\centering
	\includegraphics[width=1\textwidth]{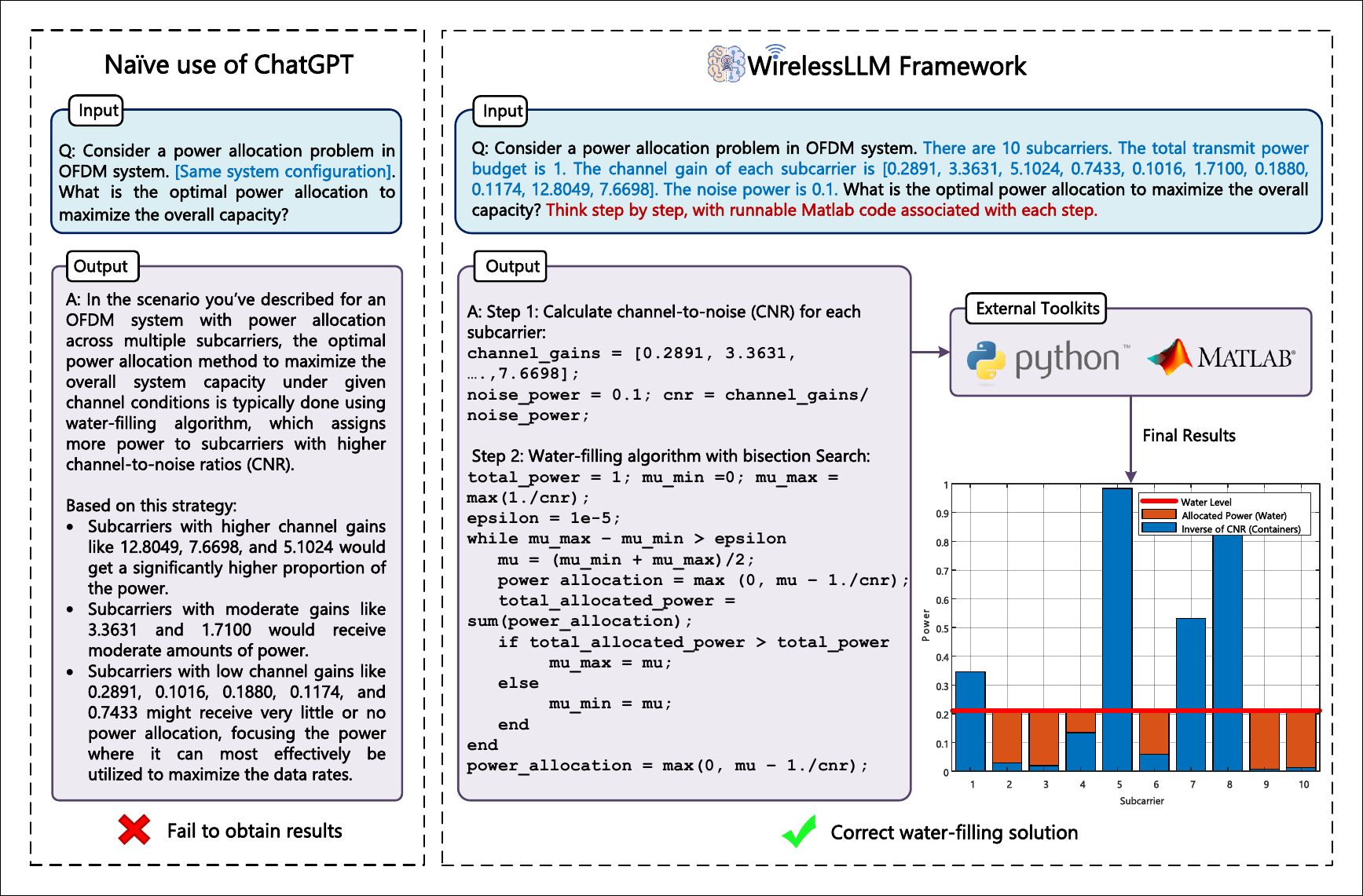}
	\caption{Comparison between ChatGPT and WirelessLLM in terms of power allocation task for OFDM systems.}
	\label{fig:LLM4Opt_Comparison}
\end{figure*}

Wireless communication systems have witnessed significant advancements over the past few decades, progressing from 1G to the current 5G, and they are envisioned to keep evolving due to the continuously emerging innovations, applications, and user demands. The profound and successful development of wireless communication systems is enabled by various advanced technologies, e.g., multi-antenna techniques, multiple access schemes, and novel network architectures, etc. Behind the advancement of integrating these driving technologies into the wireless system design, mathematical optimization plays an undoubtedly important role \cite{liu2024survey}. 

In general, the design of wireless communication systems often consists of the following three successive steps: 1) analyzing the design problem, including the objectives, requirements, implementing scenarios, etc; 2) formulating the design problem into a well-structured optimization problem; 3) solving the optimization problem with suitable and efficient algorithms and techniques.  However, as the wireless communication systems evolve substantially with more diverse design requirements, dynamic environments, user demands, etc, the optimization problems arising from the system design become increasingly complicated, e.g., from convex to non-convex problems, and from continuous to mixed-integer design variables. Consequently, solving these problems in an empirical manner, which heavily relies on the expertise of researchers or operators, becomes more challenging.

To address this challenge, we propose WirelessLLM as a universal solution for optimization problems in wireless system design, which bypasses the explicit needs for optimization expertise or even telecom knowledge. The key idea is to combine the power of CoT prompting \cite{wei2022chain} and Program-aided Language (PAL) models \cite{gao2023pal}. By employing CoT, WirelessLLM is able to understand the wireless system design task through a step-by-step analysis, which decomposes the original complex task into multiple simpler sub-tasks. By further integrating PAL models, WirelessLLM generates runnable programs associated with each sub-task. The final solution is obtained by feeding the complete program into an external toolkit, e.g., Matlab or Python interpreter. In the following, we demonstrate the capability of WirelessLLM in solving an optimization problem in wireless systems by studying a specific case: optimal power allocation in an Orthogonal Frequency Division Multiplexing (OFDM) system.

A typical power allocation task for an OFDM system is to maximize the overall capacity under given channel states. As illustrated in Figure \ref{fig:LLM4Opt_Comparison}, if we directly provide the task description and channel states as input to generic LLMs, e.g., ChatGPT, the LLMs are unlikely to generate the allocated power values, even if they possess the corresponding telecom knowledge that the water-filling algorithm should be employed for this type of optimization problem. However, by implementing our proposed WirelessLLM framework, which modifies the input with CoT and PAL models and integrates with an external toolkit, the correct water-filling solution for the optimal power allocation can be reliably obtained. This demonstrates the superior capability of WirelessLLM for solving optimization problems in wireless communication systems.

%% file: case_study_jingwen.tex
\subsection{WirelessLLM for Spectrum Sensing}
The physical (PHY) layer plays a crucial role in wireless communication systems, as it is responsible for transmitting and receiving signals over the wireless channel. This layer encompasses a range of fundamental functions, including modulation, coding, channel estimation, equalization, and signal detection. However, conventional PHY layer solutions primarily rely on task-specific models and algorithms that are tailored to specific communication scenarios and require substantial domain expertise \cite{fontaine2024towards}.
To harness the potential of LLMs for the PHY layer, we propose the WirelessLLM framework to solve various PHY layer problems by processing the time-series data \cite{lee2024integrating}. 


In the following, we consider spectrum sensing in cognitive radio networks \cite{tong2018cooperative} as a case study to demonstrate the effectiveness of WirelessLLM in handling time-series data. As shown in Figure \ref{fig:overview}, spectrum sensing involves detecting the presence or absence of a primary user (PU) signal in a given frequency band, allowing the secondary user (SU) to opportunistically access the spectrum without causing interference. Let ${H}_0$ be the hypothesis of the absence of primary signals, and ${H}_1$ be the presence of primary signals. The received signal $x(n), n = 1,2,\ldots, N, $ at the SU is given by
\begin{equation}\label{signal}
x(n) =\left\{\begin{array}{ll}
w(n), &{H}_0,\\
s(n)+w(n), &{H}_1,\\
\end{array}\right.
\end{equation}
where $w(n)$ and $s(n)$ denote the additive white noise and the primary signal, respectively.
Assume that $w(n)$ are independent and identically distributed (i.i.d.) complex Gaussian random variables with zero mean and variance $\sigma^2_n$,
while $s(n)$ follows an i.i.d. Gaussian distribution with zero mean and variance $\sigma^2_s$.
We also assume that $w(n)$ and $s(n)$ are independent of each other.
In addition, the signal-to-noise ratio (SNR) is given by $\sigma^2_s / \sigma^2_n$.

\begin{figure}[!t]
	\centering
	\includegraphics[width=0.45\textwidth]{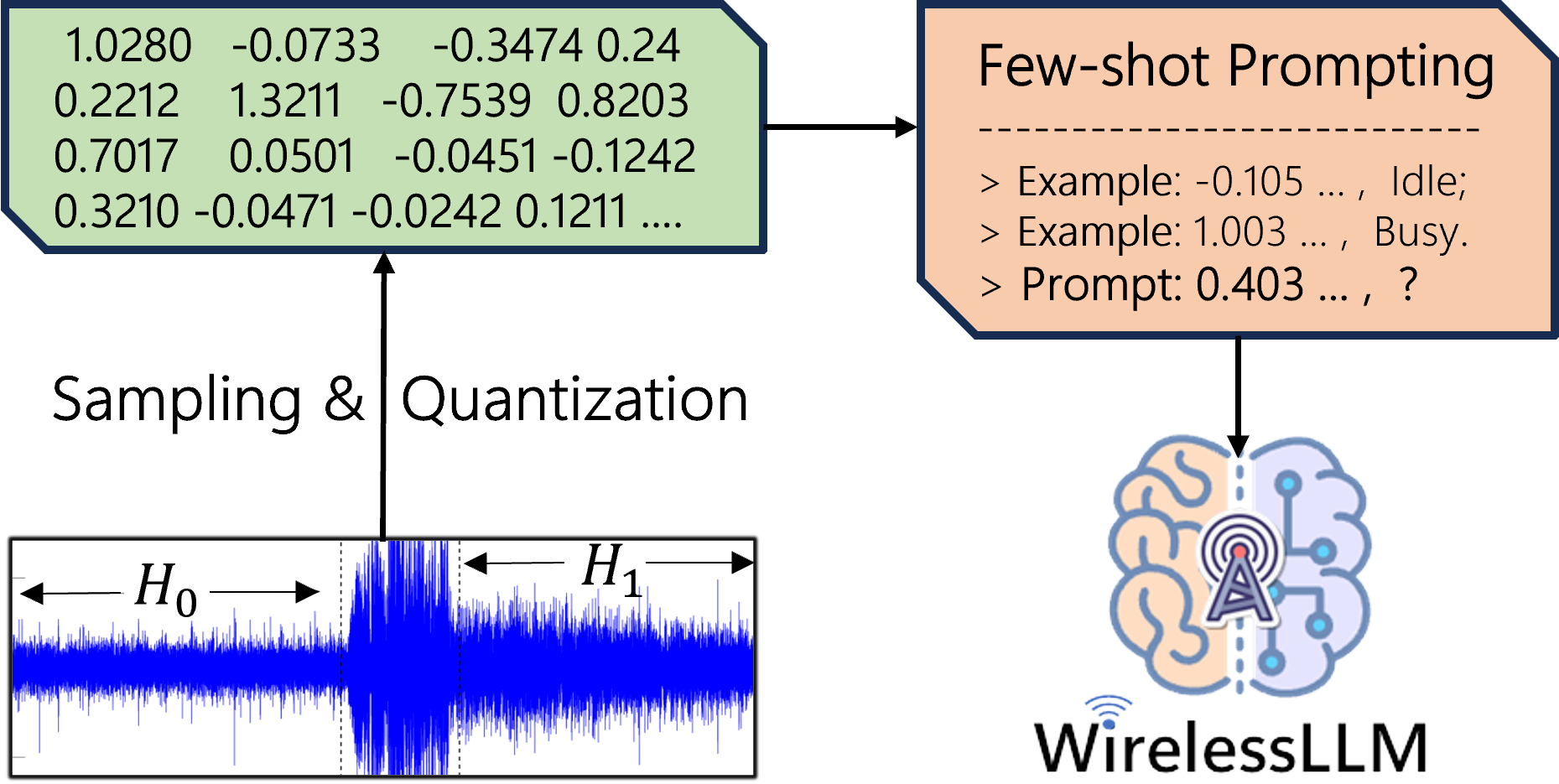}
	\caption{The WirelessLLM framework for spectrum sensing with few-shot prompting.}
	\label{LLM4PHY_Framework}
\end{figure}
We use WirelessLLM to solve this spectrum sensing problem, as depicted in Figure \ref{LLM4PHY_Framework}. In fact, this problem can be viewed as a binary classification task, where the goal is to distinguish between $H_0$ and $H_1$. WirelessLLM utilizes the few-shot prompting technique to embed domain knowledge into the LLMs, such as GPT-4 and Claude-3 Opus. Based on the provided sensing examples, the LLMs can predict the accurate spectrum states and learn the optimal classification boundary. To quantify the performance of spectrum sensing, we adopt the detection probability and false alarm rate as our performance metrics \cite{yucek2009survey}. Let $P_d=\mathrm{Pr}\{H_1|H_1\}$ and $P_f=\mathrm{Pr}\{H_1|H_0\}$ denote the detection probability and false alarm rate, respectively. In our experiments, we apply the energy detection \cite{jingwen2017} as the baseline algorithm, which serves as the optimal detector when the noise power $\sigma^2_n$ is known a prior. According to the Neyman-Pearson criterion, the threshold of the energy detection is given by \cite{jingwen2017}
\begin{equation}
\eta = \left(1+\sqrt{N^{-1}} Q^{-1} \left(P_f^{\ast}\right)\right)\sigma^2_n,    
\end{equation}
where $P_f^{\ast}$ is the target false alarm rate and $Q^{-1}(\cdot)$ is the inverse $Q$-function. Then, the spectrum sensing result is obtained by comparing the energy of the received signal $\sum_{n=1}^{N} |x(n)|^2/N$ and the threshold $\eta$.  

\begin{figure}[!t]
	\centering
	\includegraphics[width=0.45\textwidth]{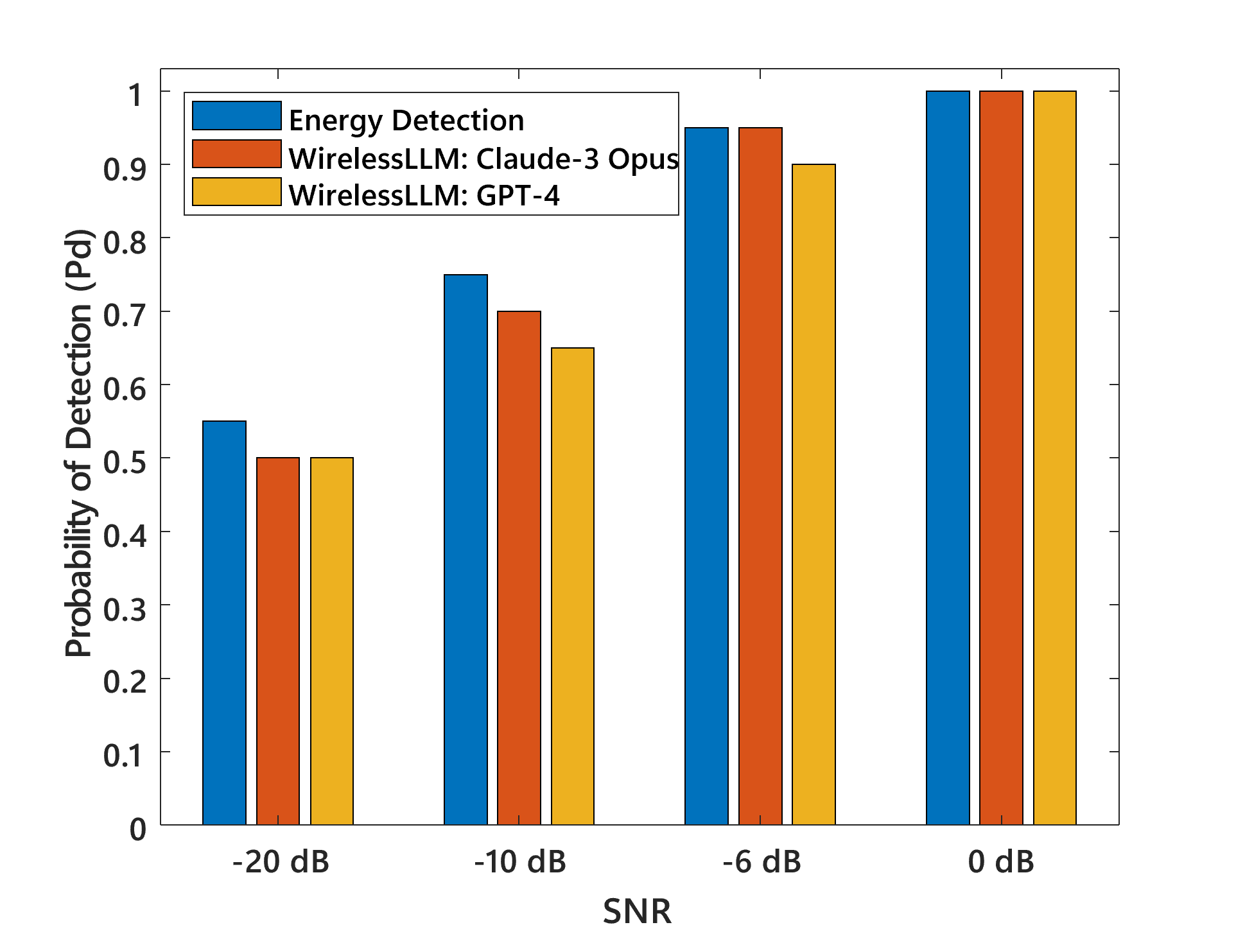}
	\caption{The detection probabilities of the energy detection, WirelessLLM with Claude-3 Opus, and WirelessLLM with GPT-4 under different SNRs.}
	\label{LLM4PHY_Results}
\end{figure}
Figure \ref{LLM4PHY_Results} illustrates the detection probabilities of energy detection, WirelessLLM with GPT-4, and WirelessLLM with Claude-3 Opus across various SNRs of $\{-20, -10, -6, 0\}$ dB. The background noise is set to $\sigma^2_n = -100$ dBm and the target false alarm rate is set to $P_f^{\ast}=0.5$. We use the APIs of GPT-4 and Claude-3 Opus to perform few-shot prompting learning. There are $20$ sensing examples, split evenly between noise and signal, with each sample containing $N=50$ individual observations. For each example, the input is the down-sampled data, as shown in Figure \ref{LLM4PHY_Framework}, and the output (or label) is the sensing result, indicating whether the spectrum corresponds to $H_0$ or $H_1$. The number of prompts used for testing is $20$. In addition, the energy detection approach is executed over $100$ independent trials. 

We see from Figure \ref{LLM4PHY_Results} that, with only a few training examples, WirelessLLM approaches the performance of the optimal energy detector, thanks to the few-shot prompting, across different SNRs. However, the WirelessLLM with Claude-3 Opus slightly outperforms the one with GPT-4 in terms of detection probability. More importantly, WirelessLLM performs equally well as the optimal energy detector when the SNR exceeds $0$ dB. In summary, by demonstrating the effectiveness of the WirelessLLM framework in the context of spectrum sensing in cognitive radio networks, we highlight the potential of this approach to revolutionize the design and optimization of PHY layer techniques in wireless communication systems.

While this case study demonstrates promising results, it is essential to further explore WirelessLLM's capabilities by delving deeper into the spectrum sensing example. One potential avenue for improvement is to employ more advanced prompt engineering techniques to design better prompts for the spectrum sensing task. Moreover, WirelessLLM's knowledge fusion and evolution capabilities can be further enhanced by incorporating a larger volume of time series data for classification tasks and fine-tuning the language model accordingly. By combining the power of LLMs with domain-specific knowledge and data, WirelessLLM has the potential to revolutionize various aspects of wireless communications, such as signal processing, channel estimation, and resource allocation.

%% file: case_study_qiong.tex
\subsection{WirelessLLM for Protocol Understanding}

\begin{table*}[t]
	\setlength{\abovecaptionskip}{-0.02cm}
	\renewcommand{\arraystretch}{1.3}
	\caption{Overview of existing datasets for network protocol understanding.}
	\label{table:Protocl_understanding_datasets}
	\centering
	\small
	\begin{tabular}{|c|c|c|c|c|c}
		\hline
		Dataset & Task & Data Sources  & Description & Open-source\\
		\hline
		TeleQnA \cite{maatouk2023teleqna}  &\multirow{4}*{\makecell{Question\\ answering}}&\makecell{Standards including 3GPP, IEEE and ITU,\\ research material and telecom lexicon}& 10,000 question-answer pairs& Yes \\
		\cline{3-5}
		\cline{1-1}
		NetEval \cite{miao2023empirical}  &~ & \makecell{Certification exam question bank, \\technical standards, etc.}& 5,269 question-answer pairs & Yes\\
		\cline{3-5}
		\cline{1-1}
		TeleQuAD \cite{holm2021bidirectional} &~&\makecell{
		3GPP, telecom data scraped from websites\\ such as ShareTechnote}&2,021 question-answer pairs&No\\
		\cline{3-5}
		\cline{1-1}
		StandardsQA \cite{roychowdhury2024unlocking}&~&3GPP and IEEE specifications&2,400 question-answer pairs&No\\
		\hline
		5GSC \cite{karim2023spec5g}&\makecell{Multi-class\\ classification}&\multirow{2}*{\makecell{Meeting minutes and technical reports\\ collected from 3GPP website}}&2,401 sentences&Yes\\
		\cline{1-2}
		\cline{4-5}
		5GSum \cite{karim2023spec5g}&Summarization &~& 713 articles & Yes\\
		\hline	
	\end{tabular}
	\vspace{-0.3cm}
\end{table*}

Wireless networks rely on a variety of communication protocols, standards, and specifications published by prominent organizations, such as the third generation partnership project (3GPP), the institute of electrical and electronics engineers (IEEE), and the international telecommunication union (ITU), to ensure reliable and efficient communications among devices \cite{bornea2024telco}. Hence, understanding the extensive protocol knowledge base plays a significant role in designing and operating wireless networks \cite{maatouk2023large}. However, keeping up with the latest protocol knowledge is time-consuming and labor-intensive due to the frequent releases and sheer volume of network protocols, which results in delays and inefficiencies in network design and management, as well as increased costs associated with manual protocol analysis. In addition, the complexity of wireless network protocols makes it challenging to identify potential issues and optimize network performance.

With the impressive capabilities in intricate language pattern capturing and complex semantic relationship understanding, LLMs offers promising opportunities to understand complicated network protocols and provide faster access to pertinent protocol information for various downstream tasks, such as network operation monitoring and network security. For example, LLMs have demonstrated proficiency in addressing complex standards-related inquiries in the wireless network domain \cite{maatouk2023teleqna}, indicating their potential to play a vital role in the design, construction, and operation of wireless networks. However, due to the limitations of LLMs, e.g., hallucinations, limited explainability, and output inconsistency, adapting the paradigm of LLMs to the wireless network domain is non-trival for unlocking the capabilities to the fullest extent for downstream wireless network tasks.

\begin{figure}[!t]
	\centering
	\includegraphics[width=0.48\textwidth]{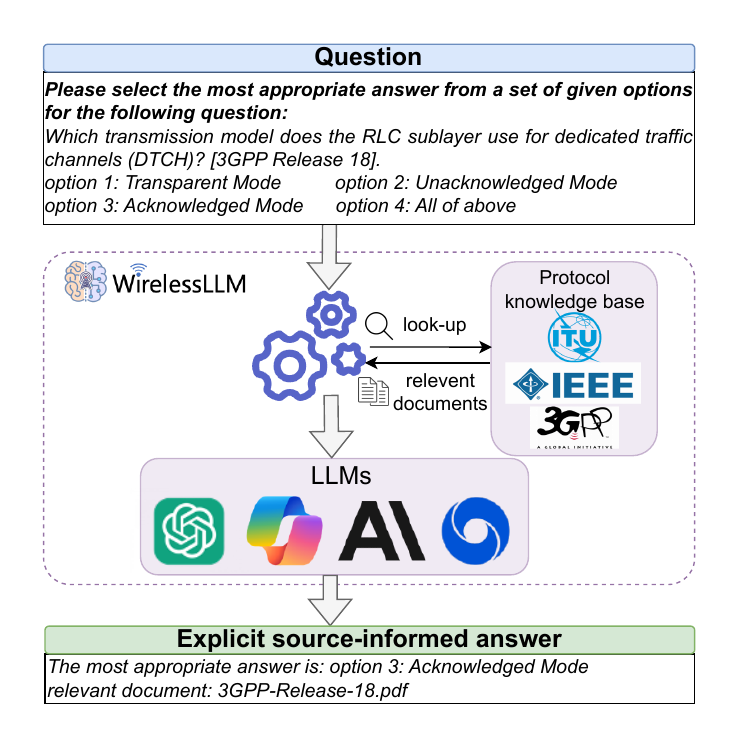}
	\caption{The WirelessLLM framework with RAG for protocol understanding.}
	\label{LLM4protocol_Framework}
\end{figure}

Considering the continuous updates of wireless network protocols, we introduce the RAG module to WirelessLLM to ensure that its output remains up-to-date and pertinent for protocol understanding. As depicted in Figure \ref{LLM4protocol_Framework}, given a question about a network protocol, WirelessLLM first scours the protocol knowledge base (e.g., 3GPP documents) to retrieve external relevant information, which is then combined with the user-entered question to construct a more comprehensive prompt as the input for LLMs. In this way, WirelessLLM can leverage the capabilities of LLMs to craft an explicit source-informed answer. By grounding the responses in real-time protocol knowledge in the wireless network domain, WirelessLLM seeks to mitigate the issue of ``hallucination" common in traditional LLMs, thereby increasing the accuracy and reliability of the generated responses. Moreover, equipped with the power of RAG, WirelessLLM can flexibly integrate with different types of LLMs (e.g., GPT-3.5, GPT-4) without the need to access their underlying model weights.

To evaluate the capabilities of WirelessLLM for network protocol understanding, several datasets of different downstream tasks in the field of wireless networks have been released recently. These datasets are summarized in Table \ref{table:Protocl_understanding_datasets}. We present a case study on the question answering task using the TeleQnA dataset, which comprises 10,000 question-answer pairs drawn from diverse sources, including standards and research articles. We maintain consistency with the data source of TeleQnA dataset and use RAG to learn information from the standards and protocols (e.g., 3GPP releases). By comparing with the widely-used LLMs (e.g., GPT-3.5 and GPT-4), we demonstrate the superior performance of WirelessLLM.

As shown in Table \ref{table:Protocl_understanding_performance}, 
WirelessLLM consistently outperforms GPT-3.5 and GPT-4 across all categories of the TeleQnA dataset by leveraging GPT-4 as the query engine. 
Notably, both GPT-3.5 and GPT-4 exhibit exceptional performance in the lexicon category. However, when faced with more intricate questions related to standards (e.g., questions in the standards specifications category), GPT-4 achieves an accuracy of $66.89\%$ while GPT-3.5 only reaches $59.05\%$, demonstrating the limitations of existing LLMs in protocol understanding. In contrast, by combining the abilities of LLMs (e.g., GPT-4) with information retrieval from the protocol knowledge base, WirelessLLM achieves an impressive accuracy of $80.34\%$ for questions related to standards specifications. This significant performance enhancement compared with GPT-3.5 and GPT-4 underscores the powerful capabilities of WirelessLLM in protocol understanding, which may further benefit a wide range of downstream wireless tasks.

\begin{table}[t]
	\setlength{\abovecaptionskip}{-0.02cm}
	\renewcommand{\arraystretch}{1.3}
	\caption{Performance comparison among GPT-3.5, GPT-4 and WirelessLLM under TeleQnA dataset.}
	\label{table:Protocl_understanding_performance}
	\centering
	\small
	\begin{tabular}{|c|c|c|c|c|c}
		\hline
		Category & GPT-3.5 & GPT-4  & WirelessLLM\\
		\hline
		Lexicon & $84.08\%$ & $89.42\%$ & $89.61\%$\\
            \hline
            Research overview & $70.45\%$ & $73.08\%$ & $73.55\%$ \\
            \hline
            Research publications & $72.38\%$ & $79.30\%$ & $79.78\%$ \\
            \hline
            Standards overview & $66.12\%$ & $76.54\%$ & $86.54\%$ \\
            \hline
            Standards specifications & $59.05\%$ & $66.89\%$ & $80.34\%$\\
		\hline	
	\end{tabular}
	\vspace{-0.3cm}
\end{table}

%% file: 6.Challenges.tex
\section{Challenges and Open Questions}\label{sec:challenges}
Although WirelessLLM presents a promising approach to enable intelligent decision-making and optimization in complex and dynamic wireless environments, realizing its full potential requires overcoming several key challenges. In this section, we delve into three critical aspects that need to be addressed: How to train WirelessLLM given data scarcity and resource constraints, how to deploy these immense models on diverse edge devices and networks, and how to safeguard the security and privacy of WirelessLLM systems. By examining these challenges and discussing potential solutions and open questions, we aim to provide a roadmap for advancing WirelessLLM research and unlocking its profound impact on next-generation wireless networks. We believe that addressing these challenges will be crucial for the integration of WirelessLLM capabilities into practical wireless systems.

\subsection{How to Train WirelessLLM?}
The training of WirelessLLM poses several challenges that need to be addressed to ensure its effectiveness and generalization across various tasks in the wireless domain. First, due to the complexity and diversity of wireless networks, as well as privacy concerns, there is a relatively small amount of wireless data available for training WirelessLLM. Moreover, the lack of annotated data for specific wireless tasks, such as decision-making, further exacerbates this challenge. Second, wireless networks can vary in terms of their topologies, protocols, and interference levels, making it difficult to train WirelessLLM that can effectively generalize to different wireless networks. Besides, as WirelessLLM is expected to achieve ``one model for all'' with superior performance and strong generalization for various wireless tasks, the design of the loss function should reflect the objectives of the tasks and the characteristics of the wireless network, which is challenging particularly when annotated data are scarce. Third, the training of WirelessLLM requires significant computational resources, limiting its scalability in resource-constrained environments. Consequently, there exists a trade-off between model complexity and computational resources for the training process of WirelessLLM.

To address these challenges, several solutions have been proposed. For instance, transfer learning can be employed in the wireless domain by fine-tuning a pre-trained LLM on a smaller wireless network dataset. Data augmentation techniques can also be used to generate synthetic data specific to the wireless domain, which can be utilized for training WirelessLLM. To tackle the challenge of designing an appropriate loss function, multi-task learning can be leveraged to train WirelessLLM on multiple related tasks simultaneously. In resource-constrained environments, model compression techniques can be leveraged to reduce the size and complexity of WirelessLLM while preserving its effectiveness. 

However, several open questions remain that need to be addressed to improve the effectiveness and generalization of WirelessLLM. First, given the limited availability of wireless data, how to effectively transfer the knowledge from pre-trained LLMs to WirelessLLM? Second, how to design loss functions that can generalize across different wireless scenarios, particularly when annotated data are limited. Moreover, there is a need for better evaluation methodologies to accurately measure the effectiveness and generalization of WirelessLLM. Lastly, the impact of WirelessLLM on the overall network performance, such as energy consumption or latency, needs to be taken into account. Addressing these open questions is crucial for training an effective and generalizable WirelessLLM that can be applied in a wide range of wireless applications.


\subsection{How to Deploy WirelessLLM?}
In practice, deploying WirelessLLM in wireless networks introduces several challenges, primarily due to the immense size of LLMs like GPT-4, the heterogeneous nature of wireless networks, and critical latency requirements for real-time applications. First, LLMs consist of billions of parameters, which poses specific challenges when considering their deployment in wireless networks. For instance, edge devices typically have limited computational power, storage, and memory, making them insufficient for directly handling such resource-intensive models. Distributing and maintaining these large models across numerous edge devices can be logistically challenging and inefficient, as it requires substantial resources and coordination. Second, wireless networks are diverse, comprising various devices connected through different network types, such as 5G, Wi-Fi, and LTE, each with varying capabilities. These networks may experience fluctuating bandwidth, which affects the ability to consistently communicate large amounts of data, including model parameters or updates. Lastly, network latency can vary significantly, impacting the performance of applications that depend on real-time interactions powered by LLMs.

However, several key aspects can be considered for the effective deployment of WirelessLLM in wireless networks. First, techniques like quantization, pruning, and knowledge distillation can be employed to reduce the size of LLMs without significantly compromising their performance, making them more suitable for edge devices with limited resources. Second, leveraging the collective computational power of edge devices through distributed computing techniques can help overcome the limitations of individual devices, enabling the efficient execution of LLMs across the network. Third, developing strategies for efficiently updating LLMs across wireless networks, considering factors such as available bandwidth, device capabilities, and application requirements, can ensure that models remain up-to-date and perform optimally. 

Moreover, the integration of LLMs into wireless communications can be greatly enhanced through the use of edge computing and federated learning. Edge computing enables the deployment of WirelessLLM at the edge of wireless networks \cite{lin2021optimizing}, which makes the LLM models closer to end devices and allows for real-time inference and optimization of wireless tasks such as power allocation, spectrum sensing, and protocol understanding. This reduces latency and bandwidth usage compared to running the models in the cloud. Federated learning, on the other hand, provides a privacy-preserving approach to continuously update and evolve the knowledge of WirelessLLM models based on data and feedback from end devices and base stations, without the need for central aggregation of raw data. Therefore, by addressing these challenges and adopting appropriate strategies, the deployment of WirelessLLM in wireless networks can be made more feasible, unlocking the potential for powerful and intelligent applications at the edge.

\subsection{How to Safeguard WirelessLLM?}
The integration of WirelessLLM into wireless networks will inevitably raise security and privacy concerns, due to the intrinsic vulnerabilities of LLMs and their reliance on a substantial volume of public data for training. The primary security concern is that LLMs are easily being attacked, either during the training process or in the deployment stage. During the training process, malicious actors can introduce harmful data to influence the behavior of LLMs, since the training datasets are typically massive and most of them are publicly available on the internet. When LLMs are deployed for use, they receive inputs from users and generate responses. This process can be easily manipulated when attackers use carefully crafted prompts to deceive the LLMs to generate unintended responses or to exploit the LLMs themselves.

In addition to the intrinsic security issue of LLMs, the privacy concerns also demand significant attention. As mentioned before, LLMs are trained on massive datasets and inadvertently memorize and regurgitate sensitive information from their training data. Once they encounter adversarial attacks, they are quite likely to reveal the sensitive information. Moreover, the complexity and lack of transparency in how LLMs process and generate responses make it difficult to control and predict what information they may disclose, thereby further obscuring privacy risks.

There are certainly possible strategies to address the above security and privacy concerns when implementing WirelessLLM. To mitigate the potential harmful training data, robust data validation and sanitization techniques can be employed to detect and remove anomalous or suspicious data before being incorporated into the training dataset. To defense WirelessLLM from detrimental attacks, we can employ adversarial training approaches to expose WirelessLLM to the adversarial examples during training or develop monitoring systems to detect and respond to unusual input patterns. As for the concern of sensitive information leakage, one simple solution is to anonymize the training data. Besides, we can apply differential privacy methods during the training process, making it more difficult to reconstruct the sensitive information. In addition, distributed learning method, such as federated learning \cite{shao2023survey}, can also be utilized to mitigate the risks of sensitive data leakage. By using federated learning, a central server is able to orchestrate multiple distributed devices to jointly train the LLMs without requiring their local data, which could enhance the privacy preservation. To improve the model transparency of WirelessLLM, we can develop explainable AI tools or white-box LLMs to provide insights into the decision-making process of WirelessLLM, thereby improving its predictability and reliability. It is important to note that even with the aforementioned techniques, safeguarding WirelessLLM remains an important and challenging task, not to mention they are not all fully reliable. Consequently, seeking and developing advanced and effective techniques to make WirelessLLM safer and more trustworthy continues to be an open problem. 